\documentclass[conference]{IEEEtran}

\usepackage{ifpdf}
\ifpdf

\ifCLASSOPTIONcompsoc
  \usepackage[nocompress]{cite}
\else
  \usepackage{cite}
\fi
\ifCLASSINFOpdf \else \fi

\usepackage{amsmath}
\usepackage{algorithmic}
\usepackage{array}
\usepackage{bm}
\usepackage[pdftex]{graphicx}
\usepackage{subfigure}
\usepackage{multirow}
\usepackage{booktabs}
\usepackage{indentfirst}
\usepackage{amsfonts,amssymb}
\usepackage{lineno}
\usepackage{color}

\renewcommand\thesection{\Roman{section}}

\begin{document}

\title{From PHY to QoE: A Parameterized Framework Design}

\author{
	\centerline{Hao Wang, Lei Ji, and Zhenxing Gao}\\
    \IEEEauthorblockA{
		Email: hunter.wanghao@huawei.com  \{jilei11, gaozhenxing\}@hisilicon.com}
}

\maketitle

\begin{abstract}
The rapid development of 5G communication technology has given birth to various real-time broadband communication services, such as augmented reality (AR), virtual reality (VR) and cloud games. Compared with traditional services, consumers tend to focus more on their subjective experience when utilizing these services. In the meantime, the problem of power consumption is particularly prominent in 5G and beyond. The traditional design of physical layer (PHY) receiver is based on maximizing spectrum efficiency or minimizing error, but this will no longer be the best after considering energy efficiency and these new-coming services. Therefore, this paper uses quality of experience (QoE) as the optimization criterion of the PHY algorithm. In order to establish the relationship between PHY and QoE, this paper models the end-to-end transmission from UE perspective and proposes a five-layer framework based on hierarchical analysis method, which includes system-level model, bitstream model, packet model, service quality model and experience quality model. Real data in 5G network is used to train the parameters of the involved models for each type of services, respectively. The results show that the PHY algorithms can be simplified in perspective of QoE.
\end{abstract}

\IEEEpeerreviewmaketitle

\section{Introduction}

With the popularization of smart terminals and the rapid development of mobile services, user behavior has undergone tremendous changes \cite{refer1}. It is no longer limited to traditional mobile services, but more immersive services, such as augmented reality (AR), virtual reality (VR), and cloud games. Considering the tremendous rate improvement, 5G radio access network (RAN) may provide a wireless mode for these services. However, more popular services are still concentrated in web browsing, video, voice, and traditional mobile gaming. Even so, the quality of these services is not entirely dependent on throughput. As user-centric services is becoming popular, studying the quality of experience (QoE) of users will be one of the core issues of future beyond 5G (B5G) wireless networks \cite{refer2}.

\begin{figure*} [htb]
\centering
\includegraphics[width=0.64\textwidth]{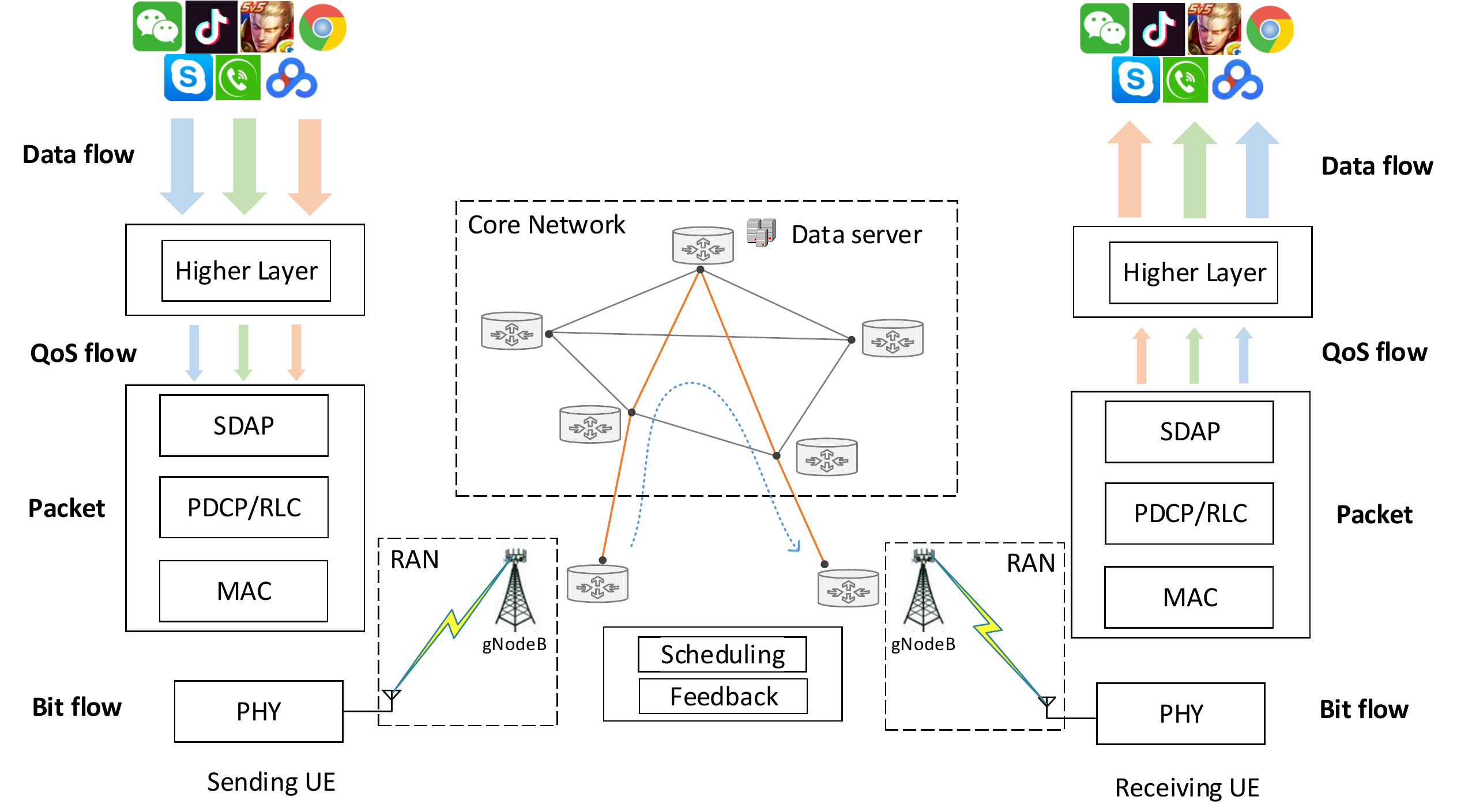}
\caption{Protocol stack of E2E data transmission.}
\label{fig}
\end{figure*}

Currently, there is no clear and strict definition of QoE in academia. In general, QoE is a measurement used to evaluate services, and it depends on the interaction between users and services\cite{refer3}, including not only objective factors such as service type and terminal parameters, but also subjective factors such as people's interest and emotion. Therefore, it is difficult to establish a comprehensive QoE assessment model. There are two well-known methods to evaluate QoE, which are subjective evaluation approach and objective evaluation approach \cite{refer4}. The subjective evaluation method is based on the testers' satisfaction of the service \cite{refer5}, namely the mean opinion score (MOS), but this method usually has high cost and poor applicability. Objective methods generally refer to the use of statistics or machine learning methods to establish parameterized models, and thus to establish a relationship between input parameters and QoE. Objective approaches have relatively high realizability. Currently, there are some relevant studies on different business types, such as literatures in \cite{refer6,refer7,refer8}. However, QoE is still an upper-layer criterion, which cannot be directly used in physical layer optimization.

Cross-layer optimization may give us some inspiration. Authors in \cite{refer9} investigated the cross-layer resource allocation in  heterogeneous wireless access network, and formulated the uplink energy-efficient video transmission into a bi-convex programming problem. Moreover, \cite{refer10} established an interference-aware cross-layer scheme for video transmission, and considered the joint physical layer (PHY) and medium access control layer (MAC) optimization. In addition, there are researches focusing on cross-layer rate adaptation mechanism \cite{refer11}, where the throughput change from PHY to application layer (APP) is analyzed. To the best of our knowledge, none of the work in literature formulated the relationship between PHY and QoE from the perspective of user equipment (UE).

Therefore, to bridge the research gap, this paper focuses on designing as follows,
\begin{itemize}
\item {A new QoE-based optimization criterion for PHY algorithm is proposed, which provides new ideas for reducing complexity and power consumption.}
\end{itemize}
\begin{itemize}
\item {Different from traditional network modeling, we formulate the end-to-end transmission from perspective of UE. A five-layer parameterized model is proposed which establishes the relationship between PHY and QoE.}
\end{itemize}
\begin{itemize}
\item {A data packet model based on the loss queuing model is proposed, which translates the physical layer performance into packet-level performance.}
\end{itemize}
\begin{itemize}
\item {This paper uses real 5G terminal test data to calibrate and verify the proposed model under several popular services.}
\end{itemize}

The reminder of this paper is organized as follows. In Section II, the end-to-end (E2E)  protocol process is briefly introduced. The proposed framework is elaborated in Section III. Simulation results and analysis are provided in Section IV. Finally, Section V concludes the whole paper.

\section{E2E Protocol Process}

Traditional optimization of PHY algorithm often considers the criterion of maximizing spectrum efficiency or minimizing error. When PHY pursues the ultimate performance, it will inevitably introduce additional complexity and power consumption. In fact, there are several mechanisms of higher protocol layer to protect the quality of service (QoS), such as automatic repeat request (ARQ) and integrity protection. On the other hand, fairness is also a non-negligible factor when base station schedules resource among multiple users. Above all, the gains brought by the physical layer algorithm may be diminished. Before introducing the framework, let us take a brief look at the E2E protocol process in 5G-RAN.

As depicted in Fig. 1, the sending UE first performs uplink transmission with its associated base station, and maps the data flow from APP into a QoS flow according to the QoS-rule issued by the user plane function (UPF). After that, through service data adaptation protocol (SDAP), packet data convergence protocol (PDCP), radio link control (RLC), and MAC layer processing, the data is stored in the buffer in the form of data packets waiting to be sent. After receiving the uplink resource grant from the base station, those data packets in buffer are processed by the PHY and transferred into a bit stream, and afterwards are transmitted to the base station through the uu air interface. Once receiving the PHY bit stream, the base station performs demodulation and decoding according to a predetermined format, and then performs the reverse unpacking process of the protocol stack. If the SDAP layer SDU information is successfully acquired, the GPRS tunnelling protocol for user-plane (GTP-U) and user datagram protocol / internet protocol (UDP/IP) encapsulation is performed, and then these packets enter the core network for routing. If necessary, the data will first be routed to data server, otherwise it is routed directly to the base station associated with the receiving UE. The base station unpacks IP/UDP, GTP-U, and then processes the data through SDAP, PDCP, RLC, and MAC layer protocols, and then the data is once again buffered in the transmission queue. After obtaining the downlink resource grant, the data is processed by the PHY and transmitted to the receiving UE through the uu air interface. After the reception, UE performs the reverse unpacking of L2, and the entire process of E2E data transmission is completed. Please refer to the relevant 3GPP protocol for detailed information, which is beyond the scope of this article.

\section{five-layer framework design}

Refering to the ideas oriented from protocol stack, we establish a multiple sub-layer parameterized framework, which includes system-level model, bitstream model, packet model, service quality model and experience quality model. These sub-models are cascade-connected as shown in Fig. 2. The overall problem can be solved step-by-step after each sub-problem.

\begin{figure*} [htb]
\centering
\includegraphics[width=0.9\textwidth]{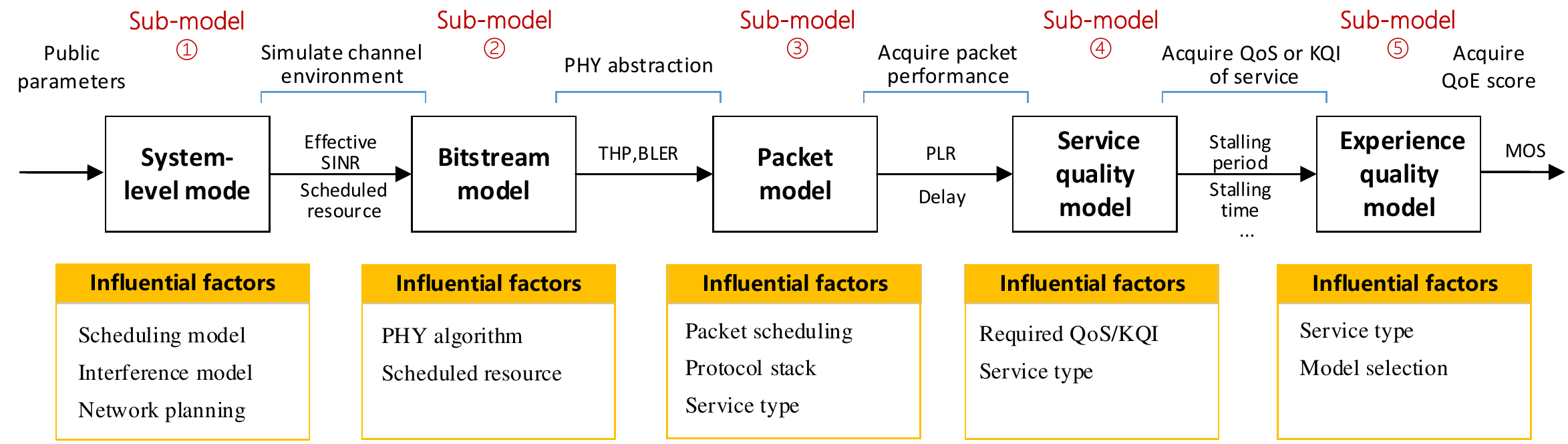}
\caption{Proposed five-layer analytical framework.}
\label{fig}
\end{figure*}

\subsection{System-level Model}

Function of system-level model is to acquire the dynamic environment of wireless channel of UE considering the scheduling in multi-cell deployment. The parameters include channel state parameter, signal to interference plus noise ratio (SINR), modulation and coding scheme (MCS), and scheduled time-frequency resources.

In fact, system modeling is a popular research topic, and there are many related works in the literature. Among which, interference model and resource scheduling model are the principle components. Currently, we adopt the closed-form analytical model based on \cite{refer12} to acquire the wanted SINR and scheduled resource in orthogonal frequency division multiple access networks.

\subsection{Bitstream Model}

Bitstream model reflects the performance of physical layer. Input parameters of this part include the output of system-level model and the parameter set of the algorithms that are adopted in PHY. Output parameters are the block error rate (BLER) and throughput of bitstream. In this paper, we use the popular PHY abstraction method depicted in \cite{refer13}. By configuring relevant PHY algorithms (mainly including channel estimation algorithms, demodulation algorithms, and decoding algorithms) and parameters, the throughput and BLER under different inputs can be obtained, namely the THP@SINR and the BLER@SINR curves.

\subsection{Packet Model}

Packet model builds the relationship between the bitstream and data packet. The input parameters of this model are the throughput and BLER of bitstream. The output of this model are the packet loss rate and the delay of packet transmission. Since the packet model only cares about the packet loss rate and packet transmission delay, we first formulate an E2E delay model.

\begin{figure*} [htb]
\centering
\includegraphics[width=0.74\textwidth]{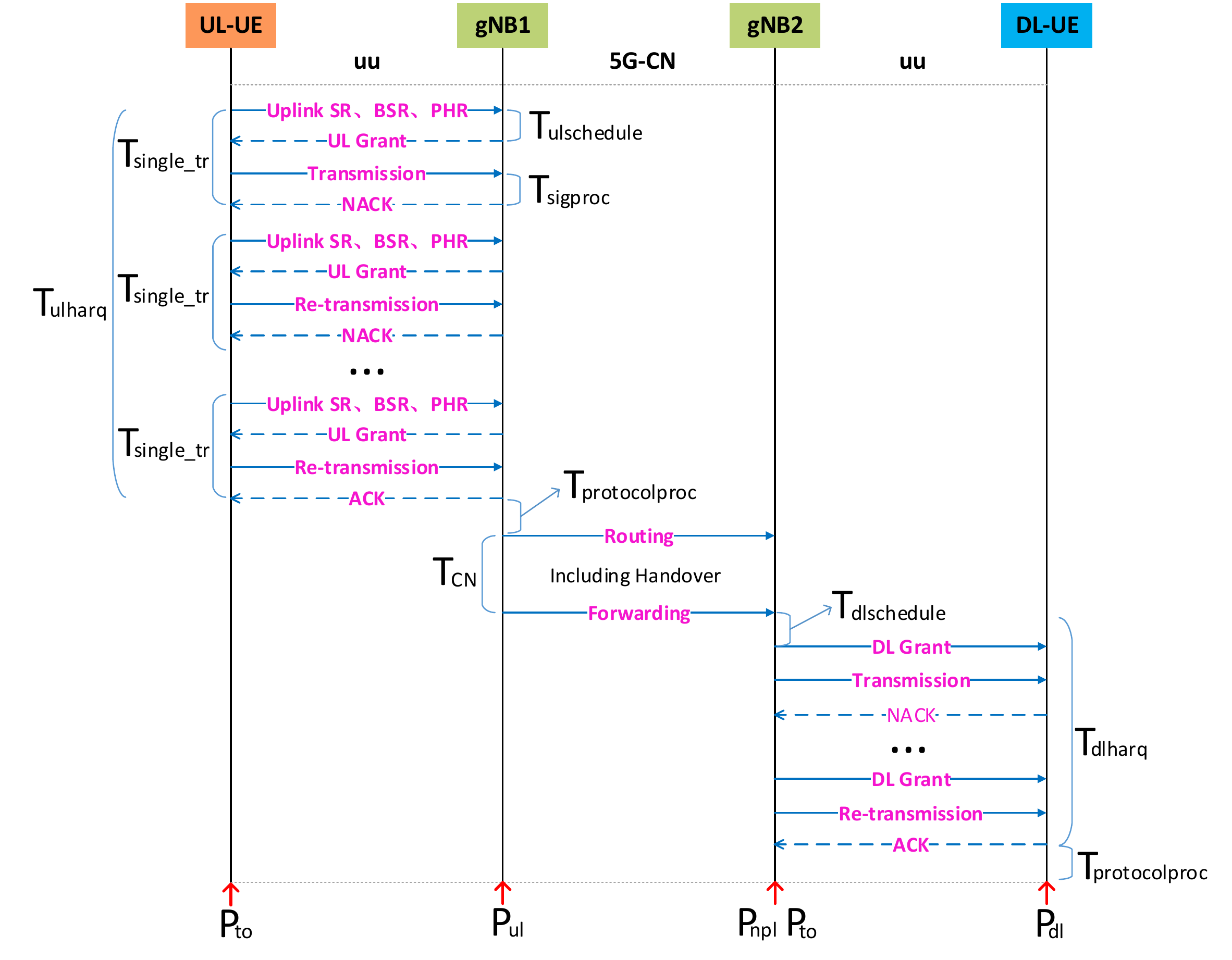}
\caption{Flow diagram of proposed packet model.}
\label{fig}
\end{figure*}

As shown in Fig. 3, the total end-to-end delay can be expressed as:
\begin{equation}
\displaystyle {T_{all} = T_{ulharq} + T_{CN} + T_{dlharq} + 2 \times T_{protocolproc}} \\
\end{equation}
where ${T_{ulharq} = E\left( N_{retr} \right) \times T_{{single}_{tr}}}$ stands for the mean time consumption of uplink HARQ procedure, and $E\left( N_{retr} \right)$ is the mean transmission times of a HARQ procedure,
\begin{equation}
\displaystyle {E\left( N_{retr} \right) = 1 + {\sum\limits_{i = 2}^{N_{harq\_ max}}{i \times {BLER}_{i - 1}}}},
\end{equation}
where ${BLER}_{i}$ represents the BLER of the i-th transmission of the particular transmission block, $i=1$ means new-transmission, $N_{harq\_ max}$ means the maximize restriction of transmission times in a HARQ procedure, and $T_{single\_ tr}$ represents the mean time of a single transmission, and
\begin{equation}
\displaystyle {T_{single\_ tr} = T_{ulschedule} + T_{sigproc} + N_{uu} \times T_{uu}},
\end{equation}
where $T_{ulschedule}$ represents the scheduling time of base station. After the base station has received the scheduling request (SR), buffer status report (BSR), and power headroom report (PHR) messages from UE, it completes the resource scheduling algorithm. The total scheduling time is related to the amount of service, business priority, and the number of involved users. $T_{ulschedule}$ can be calculated according to the average queue length and the average serving time of the TX, and the detail is shown in Eq. (\ref{T_sche}). Note that scheduling is prioritized during retransmission, and time delay can be ignored. $T_{sigproc}$ means the time consumption of signal processing, $T_{uu}$ stands for the message transmission and processing time, and $N_{uu}$ is the number of message transmission.

$T_{CN}$ is the process and routing time between base stations and core network. Note that if there is a handover during uplink transmission or downlink transmission, $T_{CN}$ will increase by $T_{ho}$,
\begin{equation}
\displaystyle {T_{CN} = T_{netrelay} + N_{ho} \times T_{ho}}
\end{equation}
where $T_{netrelay}$ is the relaying time of network, and it is related to the network topology, the degree of network congestion, and the network transmission protocol. We assume $T_{netrelay}$ as constant for simplicity.

$T_{protocolproc}$ means the protocol processing time of higher layer. $T_{dlharq}$ is similarly defined as $T_{ulharq}$, and is omitted.

\begin{figure*}[htbp]
\footnotesize
\hrulefill
\vspace*{0pt}
\begin{equation} \label{P-all}
\begin{split}
\displaystyle P_{all} = P_{to1} + \left( 1 - P_{to1} \right) P_{ul} + \left( 1 - P_{to1} - \left( 1 - P_{to1} \right) P_{ul} \right) P_{npl} +
\displaystyle \left( 1 - P_{to1} - \left( 1 - P_{to1} \right) P_{ul} - \left( 1 - P_{to1} - \left( 1 - P_{to1} \right) P_{ul} \right) P_{npl} \right) P_{to2} + \\
\displaystyle ( 1 - P_{to1} - \left( 1 - P_{to1} \right) P_{ul} - \left( 1 - P_{to1} - \left( 1 - P_{to1} \right) P_{ul} \right) P_{npl} -
\displaystyle \left( 1 - P_{to1} - \left( 1 - P_{to1} \right) P_{ul} - \left( 1 - P_{to1} - \left( 1 - P_{to1} \right) P_{ul} \right) P_{npl} \right) P_{to2} ) P_{dl}
\end{split}
\end{equation}
\hrulefill
\vspace*{0pt}
\end{figure*}

As shown in the bottom of Fig. 3, total E2E packet loss includes five component. In this paper, we assume all the components are independent to each other. As a result, the total E2E packet loss model is defined in Eq. (\ref{P-all}), where $P_{to1}$ means the packet is discarded because the transmit buffer is full. $P_{ul}$ indicates the probability that the uplink receiver fails to receive the packet correctly. $P_{npl}$ is the packet loss at the network layer, that is, the probability that a packet is correctly sent from uplink base station, but fails to reach the peer downlink base station. $P_{to2}$ and $P_{dl}$ are the corresponding loss rate of downlink and are similarly defined. Considering the fact that the value of each component is rather small, we can ignore the second order terms, expressed as
\begin{equation}
\displaystyle {P_{all} \approx P_{to1} + P_{ul} + P_{npl} + P_{to2} + P_{dl}}.
\end{equation}

Assume that the packet arrival follows the Poisson distribution with the parameter $\lambda$, and that the packet buffer follows the $M/M/1/K$ loss queuing model. Let $\mu$ represent the serving rate of 5G-RAN. Note that $\rho = \frac{\lambda}{\mu} \leq 1$ is necessary to maintain the steady state. Assuming that the maximum queue length is $K$, according to the queuing theory \cite{refer15}, the distribution of the queue length $N$ in a steady state is
\begin{equation}
\displaystyle {p_{n} = P\left\{ {N = n} \right\} = \rho^{n}p_{0},~~~n = 1,2,\ldots,K}
\end{equation}
where
\begin{equation}
\displaystyle {p_{0} = \left\{ \begin{matrix}
{\frac{1 - \rho}{1 - \rho^{K + 1}},} & {\rho \neq 1} \\
\frac{1}{K + 1} & {\rho = 1} \\
\end{matrix} \right.},
\end{equation}
and thus we have
\begin{equation}
\displaystyle {P_{to1} \approx p_{K} = \rho^{K} \cdot \frac{1 - \rho}{1 - \rho^{K + 1}}}.
\end{equation}

The average packet number $L_{avg}$ in buffer queue is defined as
\begin{equation}
\displaystyle {L_{avg} = \left\{ \begin{matrix}
{\frac{\rho}{1 - \rho} - \frac{\rho\left( {1 + K\rho^{K}} \right)}{1 - \rho^{K + 1}},} & {\rho \neq 1,} \\
\frac{K\left( K - 1 \right)}{2\left( K + 1 \right)}, & {\rho = 1.} \\
\end{matrix} \right.}
\end{equation}

Now, we can derive $T_{ulschedule}$. According to Little's law \cite{refer15}, the average waiting time $W_{avg}$ in buffer is
\begin{equation}
\label{T_sche}
\displaystyle {T_{ulschedule} \approx W_{avg} = \frac{L_{avg}}{\lambda_{e}} = \frac{L_{avg}}{\lambda\left( 1 - p_{K} \right)}},
\end{equation}
where $\lambda_{e}$ indicates the effective packet arrival strength after discarding the packets that cause the queue length larger than $K$.

The packet loss of packet transmission includes two parts. One is that the receiver fails to receive the correct data packet due to uu air interface errors, and the other is that the total delay exceeds the service tolerance window. Hence, the total packet loss rate is expressed as
\begin{equation}
\displaystyle {P_{ul} = P_{e} + \left( {1 - P_{e}} \right)P_{ul}^{loss}}.
\end{equation}

According to the queuing theory, the average serving time in the queue system $W$ follows the exponential distribution of parameter $\mu_{e} - \lambda_{e}$. Assume the service tolerance window of receiver is $T_{rxwin}$, and then the probability of packet discarding caused by exceeding the tolerance window is
\begin{equation}
\displaystyle {P_{ul}^{loss} = P\left\{ {W > T_{rxwin}} \right\} = e^{- (\mu_{e} - \lambda_{e})T_{rxwin}}}.
\end{equation}

Moreover, assuming the The tolerance window of routing in network side is $T_{netwin}$, $P_{npl}$ can be acquired as
\begin{equation}
\displaystyle {P_{npl} = ~P\left\{ {T_{CN} > T_{netwin}} \right\}}.
\end{equation}

Note that it is more reasonable if $T_{netrelay}$, $N_{ho}$ and $T_{ho}$ are random variables. However, we will study this situation in future work. $P_{to2}$ and $P_{dl}$ can be acquired similarly, and thus are omitted here.

For services like voice call, video call and mobile game, packet model shown above is compatible after setting different parameters. However, for buffered video stream service, enough buffer level is often maintained to avoid stalling event. Therefore, the major problem of this type of service is the buffer level. Packet model of buffered video stream service is formulated by utilizing the model depicted in \cite{refer14}. The model and parameters are trained using the real 5G terminal test data, the result is shown in section IV.

\subsection{Service Quality Model}

Service quality model is the interface between packet model and experience quality model. The purpose of this module is to convert the packet loss rate and transmission delay into quality indicators, so as to match the input of experience quality model. Obviously, service quality indicators are different according to the type of service. We introduce the model of video call as an example. The quality indicators include video bitrate ${Br}_{k}$, framerate $Fr_k$, definition $R_h \times R_v$, average stalling duration $T_k$, stalling times $N_k$, where $k$ is the index of video segments, and initial buffering time $T_{initial}$. $T_k$ is acquired by
\begin{equation}
\displaystyle {T_{k} = {\sum\limits_{i \in \mathcal{F}_{k}}{P_{i}^{k}*{FrameDur}_{i}}}},
\end{equation}
where $i$ is the index of video frame, ${FrameDur}_{i}$ is the duration of $i$-th frame, $P_{i}^{k}$ indicates the error probability of $i$-th frame, and we have
\begin{equation}
\displaystyle {P_{i}^{k} = {1 - \left( {1 - P_{all}^{k}} \right)^{N_{f}}}},
\end{equation}
where $N_{f}$ is the packet number in a video frame. Assuming the average packet length is $L_k$ bits, $N_{f}$ is obtained by
\begin{equation}
\displaystyle {N_{f} = \left\lceil \frac{Br_{k}}{Fr_{k} \cdot L_{k}} \right\rceil }.
\end{equation}

For services like voice call and mobile game, the service quality model are omitted because the input indicators of models described in \cite{refer6} and \cite{refer8} are packet loss rate and delay.

\subsection{Experience Quality Model}

The role of the quality of experience model is to obtain a quantitative score for the subjective experience of the terminal user. For video services, we use the parameterized QoE model described in \cite{refer7}. For voice call services such as VoNR, we use the E-model as depicted in \cite{refer6}. For mobile gaming services, we use the QoE model shown in \cite{refer8}. We can directly obtain the MOS score by sending the output of the service quality model into experience quality model. Consequently, the MOS score would ultimately be a criterion for algorithm optimization in PHY. As a result, the purpose of improving user experience or reducing power consumption while not degrading user experience may be achieved.

\section{Performance Evaluation}

In this section, we first introduce the training of the proposed framework, and then evaluate the performance.

\subsection{Model Training}

\begin{table}[!htb]
\begin{center}
\caption{Table I. Public parameter value of popular services.}
\begin{tabular}{|c|c|c|c|c|c|}
\hline
 $T_{protocolproc}$ & $T_{uu}$ & $T_{ho}$ & $T_{netrelay}$  & overhead \\
\hline
 1 ms & 1 ms & 30 ms & 5 ms & 0.95 \\

\hline
\end{tabular}
\end{center}
\end{table}

\begin{table}[!htb]
\begin{center}
\caption{Table II. Particular parameter value of popular services.}
\begin{tabular}{|c|c|c|c|c|c|}
\hline
 & $L_{k}$ & $T_{rxwin}$ & $K$ & $N_{harq\_ max}$\\
\hline
 Video call & 1000 Bytes & 300 ms & 16 & 4 \\
 \hline
 Buffered video & 1000 Bytes & 1000 ms  & 16 & 4  \\
 \hline
 Voice call & 123 Bytes & 150 ms  & 10 & 8 \\
 \hline
 Mobile game & 150 Bytes & 300 ms  & 12 & 4 \\

\hline
\end{tabular}
\end{center}
\end{table}

\begin{table}[!htb]
\begin{center}
\caption{Table III. Training results (accuracy).}
\begin{tabular}{|c|c|c|c|c|c|}
\hline
 & Training data set & Test data set \\
\hline
 Video call & 0.968 & 0.934 \\
\hline
 Buffered video & 0.964 & 0.898  \\
\hline
 Voice call & 0.977 &  0.857  \\
\hline
 Mobile game & 0.98 & 0.959 \\
\hline
\end{tabular}
\end{center}
\end{table}

We use the real test data of 5G terminal to train the parameters of proposed model. The label is the service quality data provided in test data, such as screen recording video, packet loss rate, delay parameters, etc. The training results are shown in Table I and II, where part of the parameters vary according to different services. Moreover, we summarize the accuracy of the model in both training set and test set, which are shown in Table III.

\subsection{Evaluation Results}

\begin{figure*} [!htb]
\centering
\includegraphics[width=0.98\textwidth]{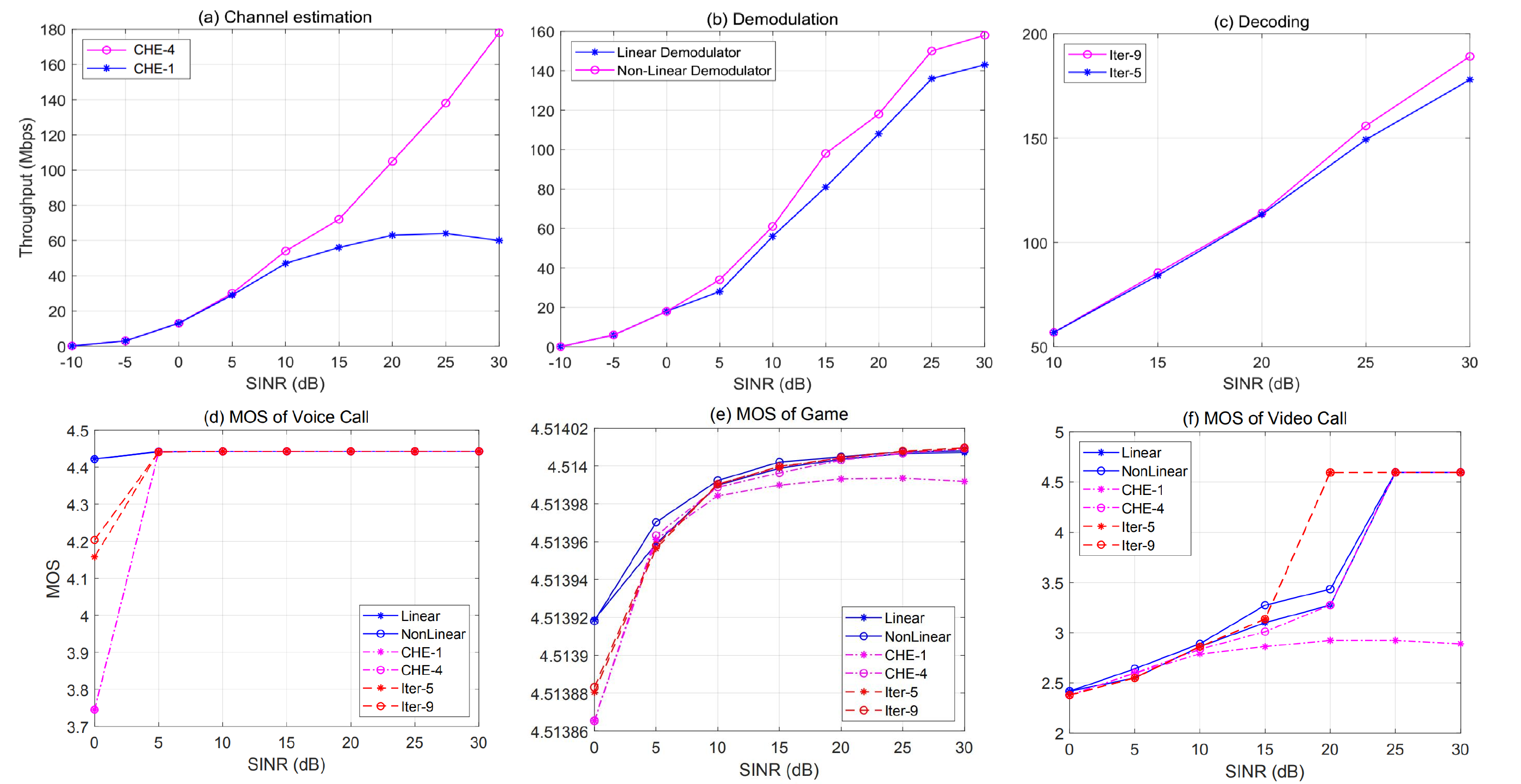}
\caption{Performance comparison of PHY algorithms.}
\label{fig}
\end{figure*}

For channel estimation algorithm \cite{refer16}, we compare the throughput performance (default BLER is 10\%) of four pilot columns and one pilot column, which are labeled as `CHE-4' and `CHE-1' in Fig. 4(a), respectively. Meanwhile, linear demodulator and non-linear demodulator \cite{refer17} are also evaluated for comparison in Fig. 4(b). Moreover, performance of decoding algorithm \cite{refer18} is verified by setting different maximum iteration number. We set `CHE-4', non-linear demodulator and `Iter-9' as baseline, respectively, and observe the MOS reduction by reducing the throughput.

We simulate the MOS of three typical services based on the throughput results in Fig. 4(a)-(c), and the results are shown in Fig. 4(d)-(f). Since video call has the highest demand for data traffic, a decrease in throughput is more likely to cause a decline in QoE. At the same time, the traffic requirements of voice call service and mobile game service are relatively small, but the excessive decrease of throughput will still lead to the increase of packet loss rate and delay. Thus, the QoE of mobile games will also worsen because mobile game is sensitive to packet loss rate and delay. Furthermore, the results also show that the change of channel estimation algorithm has a greater impact on MOS than demodulation and decoding algorithm. The reason is that using single-column pilot will lead to a large throughput (66.2\%) loss, but the simplified version of demodulation and decoding algorithm will not cause such huge throughput loss.

Above all, we can conclude that the deterioration of changing PHY algorithm is diminished in perspective of QoE, because the percentage of reduction in MOS is less than that in throughput. For example, in Fig. 4(f), the average MOS performance of non-linear demodulator is only 0.065 ( 0.27\% ) larger than linear demodulator. However, the computation complexity, namely energy consumption, of non-linear demodulator is nearly twice as lager as linear demodulator. Therefore, linear demodulator is more energy efficient than non-linear demodulator in terms of QoE in this scenario. This result proves that using MOS as PHY criterion is more energy efficient than traditional criterion, and that the PHY algorithms can be simplified in perspective of QoE.

\section{Conclusion}

In this paper, we use QoE as the optimization criterion of the physical layer algorithm. In order to establish the relationship between PHY and QoE, we propose a five-layer framework, which includes system-level model, bitstream model, packet model, service quality model and experience quality model. We use the real data in 5G network to train the parameters of the involved models for each type of services, respectively. The results show that the PHY algorithms can be simplified in perspective of QoE, which verifies the feasibility of the proposed framework.


\begin{thebibliography}{1}

\bibitem{refer1} F. Hu, et al., ``A Vision of an XR-Aided Teleoperation System toward 5G/B5G," in \emph{IEEE Communications Magazine}, vol. 59, no. 1, pp. 34-40, January 2021.

\bibitem{refer2} A. A. Barakabitze, et al., ``QoE Management of Multimedia Streaming Services in Future Networks: A Tutorial and Survey," in \emph{IEEE Communications Surveys \& Tutorials}, vol. 22, no. 1, pp. 526-565, Firstquarter 2020.

\bibitem{refer3} Y. Wang, et al., ``A Data-Driven Architecture for Personalized QoE Management in 5G Wireless Networks," in \emph{IEEE Wireless Communications}, vol. 24, no. 1, pp. 102-110, February 2017.

\bibitem{refer4} X. Liu, et al., ``KQIs-Driven QoE Anomaly Detection and Root Cause Analysis in Cellular Networks," in Proc. \emph{2019 IEEE Globecom Workshops (GC Wkshps)}, 2019, pp. 1-6.

\bibitem{refer5} Recommendation ITU-T P.800.1, Mean Opinion Score (MOS) terminology, 2006.

\bibitem{refer6} Recommendation ITU-T G.107, The E-model: a computational model for use in transmission planning, 2015.

\bibitem{refer7} Recommendation ITU-T P.1203, Parametric bitstream-based quality assessment of progressive download and adaptive audiovisual streaming services over reliable transport, 2016.

\bibitem{refer8} Recommendation ITU-T G.1072, Opinion model predicting gaming quality of experience for cloud gaming services, 2020.

\bibitem{refer9} L. Xu and W. Zhuang, ``Energy-Efficient Cross-Layer Resource Allocation for Heterogeneous Wireless Access," in \emph{IEEE Transactions on Wireless Communications}, vol. 17, no. 7, pp. 4819-4829, July 2018.

\bibitem{refer10} J. Tian, et al., ``Interference-Aware Cross-Layer Design for Distributed Video Transmission in Wireless Networks," in \emph{IEEE Transactions on Circuits and Systems for Video Technology}, vol. 26, no. 5, pp. 978-991, May 2016.

\bibitem{refer11} J. Karjee, et al., ``5G-NR Cross Layer Rate Adaptation for VoIP and Foreground/Background Applications in UE," in \emph{IEEE 3rd 5G World Forum (5GWF)}, 2020, pp. 80-85.

\bibitem{refer12} D. Parruca and J. Gross, ``Throughput Analysis of Proportional Fair Scheduling for Sparse and Ultra-Dense Interference-Limited OFDMA/LTE Networks," in \emph{IEEE Transactions on Wireless Communications}, vol. 15, no. 10, pp. 6857-6870, Oct. 2016.

\bibitem{refer13} S. Lagen, et al., ``New Radio Physical Layer Abstraction for System-Level Simulations of 5G Networks," in Proc. \emph{2020 IEEE International Conference on Communications (ICC)}, 2020, pp. 1-7.

\bibitem{refer14}  V. Burger, et al., ``A Generic Approach to Video Buffer Modeling Using Discrete-Time Analysis," in \emph{ACM Trans Multimedia Comput Commun Appl.} 14(2s), 33:1-33:23, 2018.

\bibitem{refer15} A. S. Alfa, Queueing Theory for Telecommunications: Discrete Time Modelling of a Single Node System. New York, NY, USA: Springer-Verlag, 2010.

\bibitem{refer16} M. K. Ozdemir and H. Arslan, ``Channel estimation for wireless OFDM systems," in \emph{IEEE Communications Surveys \& Tutorials}, vol. 9, no. 2, pp. 18-48, Second Quarter 2007.

\bibitem{refer17} H. Kawai, et al., ``Adaptive control of surviving symbol replica candidates in QRM-MLD for OFDM MIMO multiplexing," in \emph{IEEE Journal on Selected Areas in Communications}, vol. 24, no. 6, pp. 1130-1140, June 2006.

\bibitem{refer18} V. Savin, ``Self-corrected Min-Sum decoding of LDPC codes," in Proc. \emph{2008 IEEE International Symposium on Information Theory}, 2008, pp. 146-150

\end{thebibliography}
\end{document}